\documentclass[12pt,aps,pra,nofootinbib]{revtex4-2}

\usepackage{graphicx,amsmath,amssymb,amsfonts,bm,braket,xcolor}
\usepackage{booktabs}
\usepackage{chngcntr}
\usepackage[
colorlinks=true,
linkcolor=blue,
citecolor=blue,
urlcolor=blue,
pdfborder={0 0 0}
]{hyperref}
\usepackage{soul}

\begin{document}
	
	\title{Emergent Dissipation from Fluctuating Quantum-Network Topology}
	
	\author{Alireza Nourmandipour}
	\email{anourmandip@sirjantech.ac.ir}
	\affiliation{Department of Physics, Sirjan University of Technology, 7813733385 Sirjan, Iran}
	\author{Stefano Mancini}
	\affiliation{School of Science and Technology, University of Camerino, 62032 Camerino, Italy}
	\affiliation{INFN Sezione di Perugia, 06123 Perugia, Italy}
	
	\begin{abstract}
		We investigate how dissipation can emerge in a quantum network from temporal fluctuations of the interaction
		topology. For a network whose graph is redrawn at short time intervals, the averaged
		dynamics yields a Markovian master equation: the mean network fixes the
		coherent part and the covariance fixes the dissipative structure. The
		covariance eigenvalues set the noise strengths, while its eigenvectors
		determine the collective operators on which those strengths act. Two ensembles
		with identical mean adjacency and covariance spectra but different
		eigenvectors produce distinct dynamics despite identical purity. A physically
		realizable model yields an analytically predicted dissipative gap, giving
		quasi-protected dynamics. These results establish fluctuating connectivity as
		a resource for engineering quantum dissipation and protection.
	\end{abstract}
	
	\maketitle

	\section{Introduction}
	Dissipation and decoherence are commonly modeled through the coupling a
	quantum system to uncontrolled external degrees of freedom
	\cite{Breuer2002,Harrington2022,deVega2017}. In practice, such coupling may
	arise from a thermal bath \cite{Cattaneo2021}, intrinsic material losses \cite{Charpentier2025,Murray2021}, or from noise in
	experimental control parameters \cite{Krantz2019,Huang2025}. In engineered quantum systems, however, a distinct and less explored possibility is that the effective interaction structure itself may fluctuate: couplings can be switched on and off \cite{PitaVidal2024,Wu2018,Meng2025}, and the global connectivity can be dynamically reconfigured \cite{Valencia2025,Maskara2025}. 
	
	Quantum networks provide a natural setting in which the interaction structure can be both controlled and deliberately varied \cite{Monroe2021,Duan2010}. In such networks, information is carried by excitations hopping between adjacent nodes along the edges of a graph, and the graph topology enters directly into the Hamiltonian that governs the quantum dynamics \cite{Bose2003,Faraji2023,Honeycomb2022,Adithya2026}. When the graph is random or time-dependent, the interplay between coherent transport and topological disorder becomes nontrivial \cite{Nokkala2024,Mulken2011,Pan2026}. A particularly striking possibility is that a stochastically switching topology can generate dissipation even though each individual realization of the network evolves unitarily \cite{FarooqMancini2014}. This has been demonstrated for specific random-graph ensembles \cite{FarooqMancini2014,Cabot2018}, but the general relationship between the statistics of a fluctuating graph ensemble and
	the structure of the emergent dissipative dynamics has not been systematically established \cite{Liu2026}. This raises a specific question: does the statistical organization of the fluctuating topology determine not only how strongly quantum information is dissipated, but also which quantum degrees of freedom are exposed to that dissipation?
	
	Here we address this question by deriving the effective master equation for a network whose graph is independently redrawn at short time intervals. We show that the first two moments of the graph ensemble — its mean and its covariance — play fundamentally different roles. The mean adjacency fixes the coherent Hamiltonian, while the covariance of topology fluctuations fixes the leading dissipative structure. More importantly, the covariance eigenvalues determine the strengths of the emergent dissipative channels, whereas the covariance eigenvectors determine the collective quantum operators on which those strengths act. This distinction — noise strength versus noise geometry — is the central structural result of this paper.

	We demonstrate the mechanism with two topology ensembles that share the same mean adjacency and the same covariance spectrum but differ in covariance eigenvectors. They generate inequivalent Lindblad operators and therefore produce distinct fidelity and coherence dynamics, despite having identical total purity. This shows that global mixedness alone does not characterize topology-induced irreversibility: the same fluctuation strengths can redistribute coherence among different physical sectors. The effect persists under partial correlations away from the rank-one limit, and a physically realizable latent-variable model yields an analytically predictable dissipative gap that decreases continuously to zero at the fully correlated endpoint, demonstrating the emergence of quasi-protected quantum dynamics before that endpoint is reached. These results identify fluctuating connectivity as a resource for engineering dissipation and protected quantum sectors.

	\section{Model}
	 Consider an \(N\)-site network of two-level systems and restrict attention to its single-excitation manifold, spanned by the states \(\{\ket{j}\}_{j=1}^N\), where \(\ket j\) denotes an excitation localized at site \(j\). For an edge \(e\equiv(i,j)\) connecting sites \(i\) and \(j\), define the exchange operator
	\begin{equation}
		V_{e} \equiv \ket{i}\bra{j} + \ket{j}\bra{i},
		\label{eq:Vij}
	\end{equation}
	which is the restriction of the \(XY\) interaction \(\left(X_iX_j + Y_iY_j\right)/2\) to the
	single-excitation subspace.  The node set \( {\cal V}=\{1,\dots,N\}\) is fixed, and only the edges fluctuate.
	For a given realization, the network is described by the adjacency
	entries \(A_e\in\{0,1\}\), which we treat as independent random variables.
	The Hamiltonian for a graph realization $\cal G$ is then
	\begin{equation}
		H_{\cal G} = J\sum_{e=1}^{M} A_e V_e,
		\label{eq:H}
	\end{equation}
	where \(M = N(N-1)/2\) is the number of possible edges, \(A_e\in\{0,1\}\)
	indicates whether edge \(e\) is present in that realization, and \(J\) is
	the interaction scale \cite{FarooqMancini2014}. The edge configuration is
	independently redrawn after every interval \(\Delta t\).

	We describe the state of the network by a density operator \(\rho\) acting on the single-excitation manifold, and denote by \(\rho_0\) the initial state at \(t=0\). For a switching interval, averaging over graph realizations gives the exact	random-unitary channel
	\begin{equation}
		\Phi_{\Delta t}[\rho]
		=
		\mathbb{E}\!\left[
		e^{-iH_{\cal G}\Delta t}\rho e^{i H_{\cal G}\Delta t}
		\right].
		\label{eq:channel}
	\end{equation}
	Here  $\mathbb E$ denotes the expectation value with respect to the random graph.
	For integer $k\ge0$, corresponding to $t=k\Delta t$, the topology-averaged state is
	\begin{equation}
		\bar\rho_k
		=\Phi_{\Delta t}^{\,k}[\rho_0],
		\label{eq:averaged_state}
	\end{equation}
	where the bar denotes the ensemble average over the independently redrawn topology,	and $\Phi_{\Delta t}^{\,k}$ denotes the $k$-fold composition of the one-step channel. The central question is therefore which statistical properties of the graph	ensemble enter the effective quantum dynamics.

	To identify the relevant statistical properties, we write \(A_e = p_e + \delta A_e\),
	with \(p_e = \mathbb E[A_e]\), and define the edge-space covariance matrix
	\begin{equation}
		C_{ef}
		=
		\mathbb E[\delta A_e\delta A_f].
		\label{eq:covariance}
	\end{equation}
	The covariance matrix plays a central role in what follows: its eigenvalues
	will determine the strengths of the topology-induced dissipative channels, and
	its eigenvectors will determine the collective operators on which those
	channels act. The Hamiltonian separates as \( H_{\cal G} = \bar H + \delta H_{\cal G}\), with
	\(\bar H = J\sum_e p_e V_e\) and \(\delta H_{\cal G} = J\sum_e \delta A_e V_e\). For
	rapid switching, \(\lVert H_{\cal G}\rVert\Delta t\ll1\), expansion of
	Eq.~\eqref{eq:channel} and construction of the leading continuous generator yield
	\begin{equation}
		\mathcal L_{\mathrm{eff}}[\rho]
		=
		-i[\bar H,\rho]
		-
		\frac{J^2\Delta t}{2}
		\sum_{e,f}
		C_{ef}[V_e,[V_f,\rho]]
		+O(\Delta t^2).
		\label{eq:effective_generator}
	\end{equation}
	The derivation and regime of validity are given in the Appendix (Sec. I).
	The exact finite-step dynamics remains the completely positive map in
	Eq.~\eqref{eq:channel}; Eq.~\eqref{eq:effective_generator} is its
	leading-order generator in the small-\(\Delta t\) expansion, valid for
	\(\lVert H_{\cal G}\rVert\Delta t\ll1\). In particular, topology-induced
	dissipative rates scale as \(J^2\Delta t\sim J^2/\nu\), with switching rate
	\(\nu=1/\Delta t\), exhibiting motional narrowing in the fast-switching limit.

	%%%%%%%%%%%%%%%%%%%%%%% 	
	Because \(C\) is real, symmetric, and positive semidefinite, it admits the
	spectral decomposition
	\begin{equation}
		C_{ef}
		=
		\sum_\alpha c_\alpha O_{e\alpha}O_{f\alpha},
		\qquad
		c_\alpha\ge0,
		\label{eq:C_decomposition}
	\end{equation}
	with \(O\) an orthogonal matrix whose columns are the eigenvectors of \(C\).
	Defining the collective operators
	\begin{equation}
		W_\alpha
		=\sum_eO_{e\alpha}V_e,
		\label{eq:W}
	\end{equation}
	and using the double-commutator identity \([V,[V,\rho]]=-2\mathcal D[V]\rho\)
	for Hermitian \(V\), the effective generator takes the Lindblad form \cite{Lindblad1976,Gorini1976} 
	\begin{equation}
		\mathcal L_{\mathrm{eff}}[\rho]
		=
		-i[\bar H,\rho]
		+
		\sum_\alpha
		\Gamma_\alpha\mathcal D[W_\alpha]\rho
		+O(\Delta t^2),
		\label{eq:lindblad_form}
	\end{equation}
	where
	\begin{equation}
		\Gamma_\alpha
		=J^2\Delta t\,c_\alpha,
		\qquad
		\mathcal D[W]\rho
		=W\rho W-\frac12\{W^2,\rho\}.
		\label{eq:dissipator}
	\end{equation}
	Since the \(V_e\) are Hermitian, the collective operators \(W_\alpha\) are
	Hermitian, and the dissipator takes the simplified form shown above.
	Equations~\eqref{eq:C_decomposition}, \eqref{eq:W} and \eqref{eq:lindblad_form} separates two roles that are often
	conflated: \(c_\alpha\) set the strengths of the topology-induced noise modes,
	whereas \(O_{e\alpha}\) determine the collective operators on which those
	strengths act.
	
	%%%%%%%%%%%%%%%

	%%%%%%%%%%%%%%%%	
	
	This separation is summarized by the hierarchy
	\begin{equation*}
		\begin{array}{l}
			\text{mean topology}
			\longrightarrow
			\text{coherent Hamiltonian},
			\\[1mm]
			\text{covariance eigenvalues}
			\longrightarrow
			\text{noise strengths},
			\\[1mm]
			\text{covariance eigenvectors}
			\longrightarrow
			\text{noise geometry}.
		\end{array}
		\label{eq:hierarchy}
	\end{equation*}
	The first line follows directly from the definition of $\bar H$. 
	The second	line follows from the diagonalization of \(C\) in Eq.~\eqref{eq:C_decomposition}.	Each nonzero covariance eigenvalue \(c_\alpha\) defines one dissipative channel	\(\mathcal D[W_\alpha]\).	Thus the number of active topology-induced dissipative channels is	\(r=\operatorname{rank}C\le M\). Covariance eigenvalues therefore determine
	the strengths of these channels, while their eigenvectors determine the
	corresponding collective noise operators and hence the geometry of the
	dissipation.
	This hierarchy is the central structural result of the paper.	In particular, two ensembles can share the same $\bar H$ and the same set $\{c_\alpha\}$ while differing in the eigenvectors $O_{e\alpha}$, and hence in the physical operators $W_\alpha$ that carry those identical noise strengths.
	
	The same structure yields a sufficient criterion for topology-protected quantum sectors \cite{Zanardi1997,Lidar1998}. Let \(P\) project onto a decoherence-free subspace \(\mathcal H_{\mathrm{DFS}}\) satisfying, for every active mode (\(\Gamma_\alpha>0\)),
	\begin{equation}
		W_\alpha P = w_\alpha P,
		\label{eq:DFS_condition}
	\end{equation}
	with real \(w_\alpha\). It means that every state in the subspace is an eigenstate of \(W_\alpha\) with eigenvalue \(w_\alpha\), and suppose
	\begin{equation}
		\bar H P = P\bar H P,
		\label{eq:H_invariance}
	\end{equation}
	so that the subspace is invariant under the coherent evolution generated by \(\bar H\). Then, since each \(W_\alpha\) acts as the scalar \(w_\alpha\) on the subspace, every state supported on it satisfies \(\mathcal D[W_\alpha]\rho=0\) for all active modes. Topology-induced
	dissipation therefore vanishes on the subspace, and the remaining motion is unitary under \(\bar H\). If, in addition, \(\bar H\) acts as a scalar on the subspace, \(\bar H P = hP\) for some real \(h\), the subspace is stationary. The protected structure is thus controlled by common eigenspaces
	and degeneracies of the active covariance modes; the rank of \(C\) counts independent noise modes but does not by itself determine the dimension of a protected sector.

	\section{Covariance geometry}
	 We now provide a construction in which the covariance spectrum is identical for two ensembles while the corresponding eigenvector is rotated in edge space. Let be the \(N=2n\) sites divided into two equal communities \(A\) and \(B\), with \(A\) containing sites \(1,\ldots,n\) and \(B\) containing sites \(n+1,\ldots,N\). We specialize to an ensemble in which every possible edge has the same marginal occupation probability \(p_e = \xi\), where \(\xi\) is the single-edge occupation probability. This condition fixes the mean topology but leaves the covariance structure free. We choose \(\xi = 1/2\) so that the two ensembles defined below have identical covariance spectra. Let \(\bm1\) denote the all-ones vector in edge space and define
	\begin{equation}
		q_e=
		\begin{cases}
			+1, & e\text{ joins two sites in the same community},\\
			-1, & e\text{ joins sites in different communities}.
		\end{cases}
		\label{eq:q_definition}
	\end{equation}
	These two vectors define the two topology ensembles considered below: the global ensemble \(\mathcal E^{(G)}\) 
	uses \(\bm1\), whereas the
	community ensemble \(\mathcal E^{(C)}\) uses \(\bm q\) with components $q_e$.
	
	For the global ensemble \(\mathcal E^{(G)}\), all edges are controlled by a single
	Bernoulli variable \(X\) with \(\Pr(X=1)=\xi\), so each realization is either
	empty or complete. Its covariance matrix is
	\begin{equation}
		C^{(G)}=\frac14\bm1\bm1^T.
		\label{eq:CG}
	\end{equation}
	For the community ensemble \(\mathcal E^{(C)}\), we set \(A_e=X\) for
	intra-community edges and \(A_e=1-X\) for inter-community edges, so that
	intra- and inter-community links fluctuate oppositely. The resulting covariance
	matrix is
	\begin{equation}
		C^{(C)}=\frac14\bm q\bm q^T.
		\label{eq:CC}
	\end{equation}
	Because both ensembles have the same marginal occupation probability
	\(p_e=1/2\), their mean Hamiltonians coincide:
	\(\bar H^{(G)}=\bar H^{(C)}=(J/2)S\), with
	\(S\equiv\sum_{i<j}V_{ij}\). Both covariance matrices \eqref{eq:CG} and \eqref{eq:CC} have one nonzero eigenvalue
	\(M/4\) and \(M-1\) zero eigenvalues. Thus their covariance spectra are
	identical, while the nonzero eigenvectors are
	\(\bm1/\sqrt{M}\) and \(\bm q/\sqrt{M}\), respectively.

	Absorbing the normalization of the covariance eigenvectors into the rate, the
	collective topology-noise operators are
	\(W^{(G)}=S\) for \(\mathcal E^{(G)}\) and \(W^{(C)}=Q\) for \(\mathcal E^{(C)}\), with
	\(Q = \sum_{i<j} q_{ij} V_{ij}\). 
	Hence, to leading order,
	\begin{equation}
		\begin{aligned}
			\mathcal L^{(G)}[\rho]
			&=
			-i[\bar H,\rho]
			+\gamma_{\rm topo}\mathcal D[S]\rho,
			\\
			\mathcal L^{(C)}[\rho]
			&=
			-i[\bar H,\rho]
			+\gamma_{\rm topo}\mathcal D[Q]\rho,
		\end{aligned}
	\end{equation}
	where \(\gamma_{\rm topo}=\frac14J^2\Delta t\).
	
	Thus the two ensembles have
	identical mean topology, identical single-edge variances, and identical
	covariance spectra, yet different Lindblad operators.
	
	Define the symmetric state
	\(|+\rangle = (1/\sqrt N)\sum_{j=1}^N |j\rangle\) and the community-antisymmetric
	state
	\(|-\rangle = (1/\sqrt N)(\sum_{j\in A}|j\rangle - \sum_{j\in B}|j\rangle)\).
	For the global ensemble, \(S|+\rangle = (N-1)|+\rangle\) and
	\(S|r\rangle = -|r\rangle\) for every \(|r\rangle\) orthogonal to \(|+\rangle\),
	so the orthogonal complement of \(|+\rangle\) is an \((N-1)\)-dimensional
	protected sector. For the community ensemble,
	\(Q|-\rangle = (N-1)|-\rangle\), \(Q|+\rangle = -|+\rangle\), and
	\(Q|r\rangle = -|r\rangle\), so the protected sector is organized around
	\(|-\rangle\) instead. The distinguished noisy direction has therefore moved
	from \(|+\rangle\) to \(|-\rangle\), even though the covariance spectrum is
	unchanged.
	
	For a Hermitian noise operator \(W\) with eigenstates \(|a\rangle\) and
	eigenvalues \(w_a\),
	\begin{equation}
		\mathcal D[W](|a\rangle\langle b|) = -\frac12(w_a-w_b)^2 |a\rangle\langle b|.
		\label{eq:eigenoperator_diss}
	\end{equation}
	Thus a fixed dissipative strength need not act uniformly on different quantum
	coherences: the covariance eigenvector selects the collective operator whose
	eigenspaces determine which coherences are damped and which can be protected.
	
\begin{figure*}[t]
		\centering
		\includegraphics[width=\textwidth]{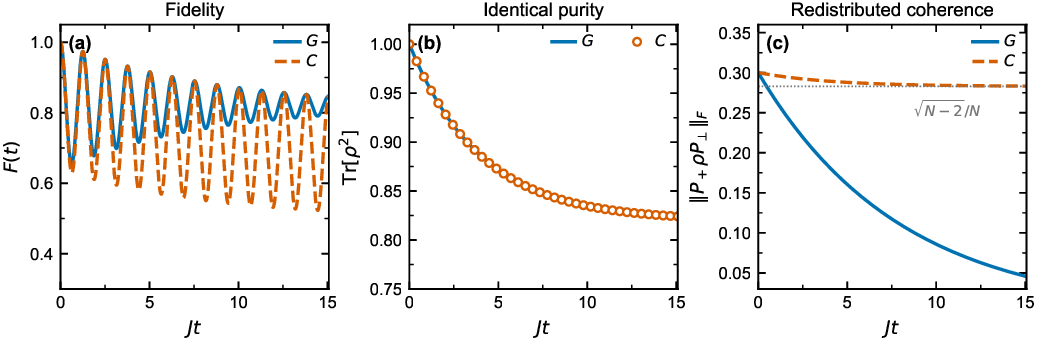}
		\caption{
			Dynamics of two topology ensembles with identical mean adjacency and
			covariance spectra but different covariance eigenvectors. (a) Fidelity,
			(b) purity, and (c) inter-sector coherence
			$\|P_{+}\rho P_{\perp}\|_{F}$. Although the total purity is identical,
			the two ensembles redistribute quantum coherence differently. Here
			$N=10$, $\xi=1/2$, $J=1$, and $\Delta t=0.01$.
		}
		\label{fig:dynamical_consequences}
	\end{figure*}

	The resulting dynamical distinction is shown in
	Fig.~\ref{fig:dynamical_consequences}. Take the initial localized state
	\(\rho(0) = |1\rangle\langle1|\), with site 1 in community \(A\). Define the
	return fidelity \(F(t) \equiv \bra1\rho(t)\ket1\) and the auxiliary quantities
	\(\omega \equiv JN/2\) and \(c \equiv \cos(\omega\Delta t)\). Because the graph is
	independently redrawn at each step, the exact finite-\(\Delta t\) evolution
	can be obtained in closed form for these two rank-one ensembles. 
	The derivation, given in the Appendix (Sec. II), yields, for the
	global ensemble
	\begin{equation}
		F^{(G)}(t)
		=
		\frac{1+(N-1)^2}{N^2}
		+
		\frac{2(N-1)}{N^2}
		c^{t/\Delta t}\cos(\omega t),
		\label{eq:fidelity_global}
	\end{equation}
	whereas for the community ensemble,
	\begin{equation}
		\begin{aligned}
			F^{(C)}(t)
			={}&
			\frac{(N-2)^2+2}{N^2}
			+
			\frac{2}{N^2}c^{t/\Delta t}\cos(\omega t)
			\\
			&+
			\frac{2(N-2)}{N^2}\cos(\omega t)
			+
			\frac{2(N-2)}{N^2}c^{t/\Delta t}.
		\end{aligned}
		\label{eq:fidelity_community}
	\end{equation}
	Thus identical mean topology and identical covariance spectra do not imply
	identical fidelity dynamics --- a direct manifestation of the
	eigenvector-controlled dissipation.
	
	%%%%%%%%%%%%%%%%%

	The sector-resolved coherence gives an even sharper distinction. With
	\(P_+ = |+\rangle\langle+|\) and \(P_\perp = I - P_+\), define
	\begin{equation}
		\mathcal C_{+\perp}(t)
		\equiv
		\left\|P_+\rho(t)P_\perp\right\|_F,
		\label{eq:sector_coherence}
	\end{equation}
	which measures the Frobenius norm of the coherence block connecting the
	symmetric sector to its orthogonal complement \cite{Yao2016}. Then
	\begin{equation}
		\mathcal C^{(G)}_{+\perp}(t)
		=
		\frac{\sqrt{N-1}}{N}|c|^{t/\Delta t},
		\label{eq:coherence_global}
	\end{equation}
	whereas
	\begin{equation}
		\mathcal C^{(C)}_{+\perp}(t)
		=
		\frac1N\sqrt{N-2+c^{2t/\Delta t}}.
		\label{eq:coherence_community}
	\end{equation}
	Thus the coherence decays to zero for the global ensemble but plateaus at a
	nonzero value \(\sqrt{N-2}/N\) for the community ensemble, even though the two
	ensembles share the same covariance spectrum.
	
	Remarkably, the total purity is identical for the two ensembles:
	\begin{equation}
		\begin{aligned}
			\operatorname{Tr}\!\left[(\rho^{(G)}(t))^2\right]
			=&
			\operatorname{Tr}\!\left[(\rho^{(C)}(t))^2\right] \\
			=&\frac{1+(N-1)^2}{N^2}
			+
			\frac{2(N-1)}{N^2}c^{2t/\Delta t}.
		\end{aligned}
		\label{eq:purity}
	\end{equation}
	Hence the eigenvector effect is not simply a difference in the amount of
	mixedness. Rather, the same fluctuation strengths — the identical covariance
	spectra — redistribute coherence among different physical quantum sectors, as
	seen directly in Fig.~\ref{fig:dynamical_consequences}.

	For \(\Delta t\ll J^{-1}\), we have \(\ln c = -\omega^2\Delta t^2/2 + O(\Delta t^4)\), so that \(c^{t/\Delta t}\to e^{-\gamma_{+\perp} t}\) with
	\(\gamma_{+\perp} \equiv \omega^2\Delta t/2 = J^2 N^2 \Delta t/8\). Here \(\gamma_{+\perp}\) is the decay rate of the sector coherence \(\mathcal C_{+\perp}(t)\). Thus, in the rapid-switching regime, the exact discrete dynamics is approximated by the effective master-equation dynamics to second order in \(\Delta t\). Numerical convergence to the effective generator follows approximately the expected second-order state-level scaling in \(\Delta t\); details are given in the Appendix (Sec. II).
	
	\textit{Partial correlations}--- The geometry effect persists continuously away from the perfectly correlated limit. To make this interpolation explicit, define the interpolated one-step channels \(\Phi^{(G)}(\eta) = \eta\Phi^{(G)}_{\rm corr} + (1-\eta)\Phi_{\rm Gilb}\) and \(\Phi^{(C)}(\eta) = \eta\Phi^{(C)}_{\rm corr} + (1-\eta)\Phi_{\rm Gilb}\), with \(0\le\eta\le1\). Here \(\Phi_{\rm Gilb}\) denotes the channel obtained from independently sampled Gilbert graphs \cite{Gilbert1959} with edge probability \(\xi=1/2\), while \(\Phi^{(G)}_{\rm corr}\) and \(\Phi^{(C)}_{\rm corr}\) denote the globally correlated and community-correlated rank-one ensembles defined above. Thus \(\eta\) is a channel-mixture parameter, not a conventional pairwise correlation coefficient. A physical realization of partial correlations is given by the latent-variable construction introduced below.

	Because all constituent ensembles have the same edge mean, both interpolations retain \(p_e=1/2\). Their covariance matrices are \(C^{(G)}(\eta)=
	\frac14[(1-\eta)I_M+\eta\bm1\bm1^T]\) and \(C^{(C)}(\eta)=\frac14[(1-\eta)I_M+\eta\bm q\bm q^T]\). The two spectra coincide for every \(\eta\), with leading eigenvalue \(c_1(\eta) = [1+\eta(M-1)]/4\) and transverse eigenvalues \(c_{\alpha>1}(\eta) = (1-\eta)/4\), while the leading collective eigenvectors remain \(\bm1\) and \(\bm q\), respectively. The difference between the corresponding effective generators is therefore
	\begin{equation}
		\mathcal L^{(G)}(\eta)-\mathcal L^{(C)}(\eta)
		=
		\frac{\eta J^2\Delta t}{4}
		\left[\mathcal D[S]-\mathcal D[Q]\right]
		+O(\Delta t^2).
		\label{eq:generator_difference}
	\end{equation}
	The leading and transverse covariance modes contribute with opposite signs; using the completeness of the covariance eigenvectors, their sum reduces to the simple form above. Thus the geometry-dependent distinction turns on continuously with the correlated component.
	
\begin{figure*}[t]
		\centering
		\includegraphics[width=\textwidth]{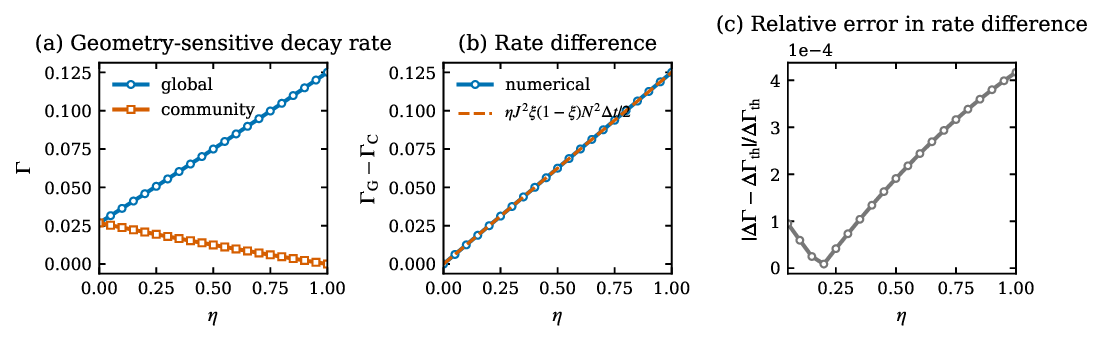}
		\caption{
			Geometry-dependent dissipation under partial topology correlations.
			(a) Decay rates of the geometry-sensitive coherence for globally
			correlated and community-correlated ensembles as a function of the
			channel-mixture parameter $\eta$. (b) Their difference compared with the
			leading-order prediction $\eta J^{2}\xi(1-\xi)N^{2}\Delta t/2$.
			(c) Relative error in the rate difference with respect to the
			rapid-switching prediction. The single curve quantifies the agreement
			shown in (b); the pronounced minimum at intermediate $\eta$ reflects
			the competition between the normalization by a prediction that
			vanishes at $\eta=0$ (which inflates the relative error at small
			$\eta$) and finite-$\Delta t$ corrections (which grow with $\eta$). Here
			$N=10$, $\xi=1/2$, $J=1$, and $\Delta t=0.01$. 
		}
		\label{fig:partial_correlations}
	\end{figure*}

	This continuous behavior is demonstrated in Fig.~\ref{fig:partial_correlations}. To quantify the geometry-dependent dissipation, consider the decay rate extracted from the coherence between the symmetric state and a residual state orthogonal to the corresponding collective direction (that is, \(\ket{+}\) for the global ensemble and \(\ket{-}\) for the community ensemble). For this coherence, the eigenvalue differences under \(S\) and \(Q\) are \(N\) and \(0\), respectively, so the rate difference obeys, in the rapid-switching regime,
	\begin{equation}
		\Delta\Gamma_{\rm th} \equiv \Gamma^{(G)}-\Gamma^{(C)}
		=
		\frac{\eta J^2\xi(1-\xi)N^2\Delta t}{2}
		+O(\Delta t^2).
		\label{eq:rate_difference}
	\end{equation}
	The numerical rates follow this linear dependence on \(\eta\), demonstrating
	that the eigenvector-dependent distinction survives continuously away from the
	rank-one endpoint.

	\section{Multimode realization}
	 The preceding examples isolate the eigenvector
	effect using rank-one covariance structures. We next ask whether partial topology
	correlations can instead generate a continuous family of covariance structures
	interpolating between the rank-one and independent-edge limits, with a
	correspondingly tunable dissipative gap.
	
	Consider again \(N=2n\) sites divided into two equal communities and introduce a
	binary latent variable \(Z=\pm1\), with
	\begin{equation}
		\Pr(Z=+1)=\Pr(Z=-1)=\frac12.
		\label{eq:Z_distribution}
	\end{equation}
	Conditional on \(Z\), distinct edges are sampled independently according to
	\begin{equation}
		\Pr(A_e=1\mid Z)=\xi+a q_e Z,
		\label{eq:latent_probability}
	\end{equation}
	where \(q_e\) is given by Eq.~\eqref{eq:q_definition}, the parameter \(a\)
	satisfies \(|a|\le\min(\xi,1-\xi)\) so that the conditional probability remains
	in \([0,1]\) for all realizations, and edge correlations are generated entirely
	by the shared latent variable \cite{Goldenberg2010}. We specialize to
	\(\xi=1/2\), for which the physical range reduces to \(|a|\le1/2\). At \(a=0\)
	the model reduces to the Gilbert ensemble, while at the endpoint \(a=1/2\) it
	reduces to the community ensemble, providing a physically realizable
	interpolation between these two limits.
	Averaging over \(Z\) leaves the marginal occupation unchanged,
	\(p_e=\xi\), while \(\operatorname{Cov}(A_e,A_f)=a^2q_eq_f\) for \(e\neq f\)
	and \(\operatorname{Var}(A_e)=\xi(1-\xi)\). Notably, the single-edge variance is
	independent of \(a\), so the diagonal of the covariance matrix is unchanged as
	the correlations are turned on.
	
	The full covariance matrix is therefore
	\begin{equation}
		C(a)
		=
		\left[\xi(1-\xi)-a^2\right]I_M
		+a^2\bm q\bm q^T.
		\label{eq:latent_covariance}
	\end{equation}
	Its spectrum contains one collective eigenvalue
	\(c_{\rm coll}(a)=\xi(1-\xi)+a^2(M-1)\) and \(M-1\) degenerate transverse
	eigenvalues \(c_{\rm loc}(a)=\xi(1-\xi)-a^2\). For \(\xi=1/2\) and \(0\le a\le1/2\), introducing the dimensionless
	correlation parameter
	\(\kappa=a/\sqrt{\xi(1-\xi)}\), with \(0\le\kappa\le1\),
	gives \(C(\kappa)=\xi(1-\xi)[(1-\kappa^2)I_M+\kappa^2\bm q\bm q^T]\), with
	\(c_{\rm coll}(\kappa)=\xi(1-\xi)[1+\kappa^2(M-1)]\) and
	\(c_{\rm loc}(\kappa)=\xi(1-\xi)(1-\kappa^2)\). Thus increasing \(\kappa\)
	transfers covariance weight into one collective mode while suppressing the
	remaining \(M-1\) modes.
	
	To quantify the resulting long-lived dynamics, let \(R_\mu\) be right
	eigenoperators of the effective Liouvillian \cite{Gravina2023},
	\(\mathcal L_{\rm eff}R_\mu=\Lambda_\mu R_\mu\), with decay rates
	\(-\operatorname{Re}\Lambda_\mu\). We define the dissipative gap as the smallest
	nonnegative decay rate among nonstationary modes,
	\begin{equation}
		\gamma_{\rm gap}
		=
		\min_{\Lambda_\mu\neq0}
		\left[-\operatorname{Re}\Lambda_\mu\right].
		\label{eq:dissipative_gap}
	\end{equation}
	Thus \(\gamma_{\rm gap}=0\) whenever the Liouvillian develops a nonstationary
	mode with purely imaginary eigenvalue.
	
\begin{figure*}[t]
		\centering
		\includegraphics[width=\textwidth]{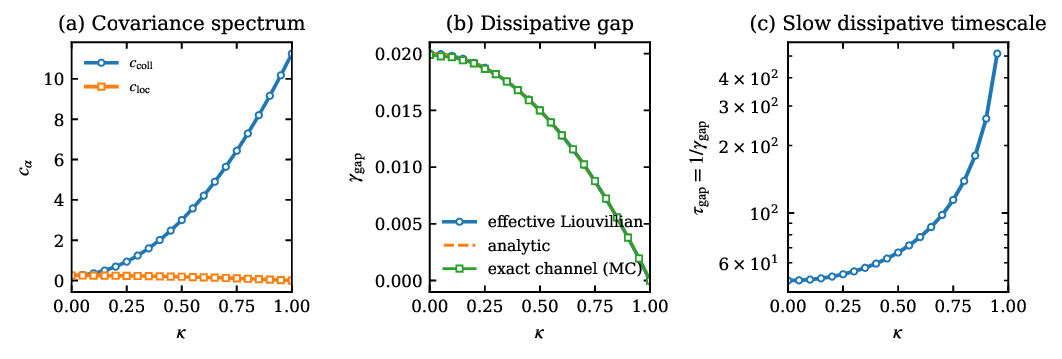}
		\caption{
			Continuous emergence of quasi-protection in a physically realizable
			partially correlated random network. (a) Collective and transverse
			covariance eigenvalues as functions of the correlation parameter $\kappa$.
			(b) Dissipative gap of the effective Liouvillian, its analytic prediction
			$\gamma_{\rm gap}=J^2\Delta t\xi(1-\xi)(N-2)(1-\kappa^2)$, and a Monte-Carlo estimate of the exact one-step random-unitary channel.
			(c) Slow dissipative timescale of the effective Liouvillian,
			\(\tau_{\rm gap}=1/\gamma_{\rm gap}\).
			Here $N=10$, $\xi=1/2$, $J=1$, and $\Delta t=0.01$.
		}
		\label{fig:liouvillian_gap}
	\end{figure*}

	The gap can be identified analytically throughout the rapid-switching regime.
	Introducing the projectors \(P_+=|+\rangle\langle+|\) and
	\(P_-=|-\rangle\langle-|\), consider the operator
	\begin{equation}
		R_{\rm slow}=P_+-P_-.
		\label{eq:slow_mode}
	\end{equation}
	Since both \(\ket{+}\) and \(\ket{-}\) are eigenstates of \(\bar H\),
	\([\bar H,R_{\rm slow}]=0\), while both projectors are diagonal in the
	\(Q\)-eigenbasis and hence \(\mathcal D[Q]R_{\rm slow}=0\). The remaining
	\(M-1\) transverse covariance modes, with collective operators
	\(W_\mu=\sum_e O_{\mu e}V_e\) orthogonal to \(\bm q\), satisfy
	\begin{equation}
		\sum_\mu\mathcal D[W_\mu]R_{\rm slow}
		=
		-(N-2)R_{\rm slow},
		\label{eq:transverse_sum}
	\end{equation}
	as follows from the explicit form of the exchange operators \(V_e\) (see
	Appendix -- Sec. III). Therefore
	\begin{equation}
		\begin{aligned}
			\mathcal L_{\rm eff}R_{\rm slow}
			=&
			-\gamma_{\rm slow}R_{\rm slow},\\
			\qquad
			\gamma_{\rm slow}
			=&
			J^2\Delta t\,\xi(1-\xi)(N-2)(1-\kappa^2).
		\end{aligned}
		\label{eq:analytic_gap}
	\end{equation}
	For the parameters used in Fig.~\ref{fig:liouvillian_gap}, this reduces to
	\(\gamma_{\rm slow}=0.02(1-\kappa^2)\).
	
	At the fully correlated endpoint \(\kappa=1\), all transverse covariance modes
	are suppressed. The kernel of the dissipator \(\mathcal D[Q]\), consisting of
	operators block-diagonal with respect to the \(Q\)-eigenspaces, has dimension
	\(\dim\ker\mathcal D[Q]=1+(N-1)^2=82\) for \(N=10\). The effective Liouvillian,
	however, annihilates only the subset of these operators that also commute with
	\(\bar H\); this subset has dimension
	\(\dim\ker\mathcal L_{\rm eff}=(N-2)^2+2=66\) for \(N=10\) and constitutes the
	space of stationary operators. The remaining \(2(N-2)=16\) operators are
	coherences between the symmetric state \(\ket{+}\) and the
	\((N-2)\)-dimensional subspace orthogonal to both \(\ket{+}\) and \(\ket{-}\).
	These operators lie in the kernel of \(\mathcal D[Q]\), since both sectors carry
	the same \(Q\)-eigenvalue \(-1\), but they do not commute with \(\bar H\). For
	each of the \(N-2\) residual states, their Liouvillian eigenvalues form a purely
	imaginary conjugate pair \(\Lambda_\pm=\pm iJN/2\), so these modes oscillate
	without dissipative decay. The zero-real-part sector of the Liouvillian
	therefore has dimension \(1+(N-1)^2=82\), containing \(66\) stationary modes
	and \(16\) nonstationary oscillatory modes. The dissipative gap closes
	continuously as \(\kappa\to1\), because the transverse covariance strength
	\(c_{\rm loc}\propto1-\kappa^2\) tends to zero and the decay rate of
	\(R_{\rm slow}\) vanishes.

	For the exact one-step channel, eigenoperators satisfy
	\(\Phi_{\Delta t}R_\mu=\mu_\mu R_\mu\), with discrete decay rate
	\(\gamma_\mu^{\rm exact}=-(1/\Delta t)\ln|\mu_\mu|\). For \(0\le\kappa<1\),
	we evaluate the ensemble average by Monte Carlo sampling of graph
	realizations, while each unitary evolution is computed exactly; at
	\(\kappa=1\), the two graph realizations are treated exactly. This provides
	a finite-switching benchmark for the effective Liouvillian gap.
	
	The effective Liouvillian for the latent-variable model takes the multimode
	form of Eq.~\eqref{eq:lindblad_form}, with the collective mode associated with
	\(\bm q\) and the remaining \(M-1\) modes orthogonal to it. At \(\kappa=0\), the
	graph reduces to independent-edge fluctuations. As \(\kappa\to1\),
	\(c_{\rm loc}\to0\), so the transverse topology-noise modes are progressively
	suppressed and the stationary manifold enlarges at the rank-one endpoint.
	
	The resulting spectral flow is shown in Fig.~\ref{fig:liouvillian_gap}. The numerically extracted gap of the effective Liouvillian follows the
	analytic prediction in Eq.~\eqref{eq:analytic_gap} throughout the parameter
	range shown, establishing that \(R_{\rm slow}\) is the slowest decaying mode
	in this regime. The Monte-Carlo evaluation of the exact one-step
	random-unitary channel provides a finite-switching benchmark and closely
	tracks the same gap. The associated slow timescale
	\(\tau_{\rm gap}=1/\gamma_{\rm gap}\) therefore grows as \(\kappa\)
	approaches unity, signaling the emergence of quasi-protected dynamics before
	the fully correlated endpoint is reached.

	\section{Discussion}
	 We have shown that the dissipative structure of a
	quantum network with stochastically switching topology is fixed by the
	covariance of the graph ensemble, and that its eigenvalues and eigenvectors
	play fundamentally different roles. The eigenvalues set the strengths of the
	topology-induced noise modes, while the eigenvectors determine the collective
	quantum operators on which those strengths act. This separation is the central
	structural result of the present work, and it implies that two network
	ensembles can share the same mean topology, the same single-edge variances, and
	the same covariance spectrum while producing inequivalent quantum dynamics.
	
	We demonstrated this explicitly with two rank-one ensembles that differ only in
	the orientation of their covariance eigenvector. The two ensembles exhibit
	different fidelity and coherence dynamics but identical total purity, showing
	that mixedness alone does not characterize topology-induced irreversibility.
	The distinction persists under partial correlations, and a physically
	realizable latent-variable model yields a continuously narrowing dissipative
	gap, with quasi-protected dynamics emerging before the fully correlated
	endpoint. These results identify fluctuating connectivity as a resource for
	engineering dissipation, coherence redistribution, and protected quantum
	sectors.
	
	The framework could in principle be implemented in programmable quantum simulators \cite{Yan2018,Monroe2021,Kyprianidis2024}, where
	a global control field provides the latent variable and local control encodes
	the community structure, so that the predicted narrowing of the dissipative gap
	can be probed through the decay of a single collective coherence. In particular, the
	coherence between the symmetric and antisymmetric collective states should
	decay at the analytically predicted rate
	\(\gamma_{\rm slow}=J^2\Delta t\,\xi(1-\xi)(N-2)(1-\kappa^2)\), providing a
	direct experimental signature of the covariance-geometry effect. A natural
	extension is to temporally correlated topology, where the instantaneous
	covariance is replaced by a two-time correlation function and the effective
	dynamics acquires memory \cite{Chisholm2021,Ding2026}. More generally, the covariance-geometry principle
	established here suggests that the statistical organization of a fluctuating
	environment, not only its strength, can be engineered to control which quantum
	degrees of freedom are protected and which are lost \cite{Kapit2017,Nallamothula2026,Bittner2026}.

\clearpage
\appendix

% Prevent section-based reset of the equation counter
\counterwithout{equation}{section}

% Restart equation numbering at A1 and format as A1, A2, A3, ...
\setcounter{equation}{0}
\renewcommand{\theequation}{A\arabic{equation}}

\section{Random-unitary channel and effective master equation}
In this Section we build the exact random-unitary channel generating the rapid-switching master equation, then diagonalize the graph covariance to disentangle topology-induced noise strengths from operator geometry.

\subsection{Rapid-switching expansion}

We consider the random-unitary one-step channel introduced in
Eq.~(3) of the main text,
% [MAIN-EQ] verify equation number of the channel in the main text
\begin{equation}
	\Phi_{\Delta t}[\rho]
	=
	\mathbb E\!\left[
	U_{\cal G}\rho U_{\cal G}^\dagger
	\right],
	\qquad
	U_{\cal G}=e^{-i H_{\cal G}\Delta t},
	\label{sm:channel}
\end{equation}
where the node set is fixed and the randomness is entirely in the edge
occupations $A_e$. The Hamiltonian is
\begin{equation}
	H_{\cal G}=J\sum_e A_eV_e=\bar H+\delta H_{\cal G},
	\label{sm:ham}
\end{equation}
with
\begin{equation}
	\bar H=J\sum_ep_eV_e,
	\qquad
	\delta H_{\cal G}=J\sum_e\delta A_eV_e,
	\qquad
	\mathbb E[\delta A_e]=0.
	\label{sm:hdecomp}
\end{equation}
For $V_e$ as in Eq.~(3) of the main text, this corresponds to the class of stochastically switching $XY$ networks introduced in
Ref.~\cite{FarooqMancini2014}.

When $\lVert H_{\cal G}\rVert\Delta t\ll1$, the unitary expansion gives
\begin{align}
	U_{\cal G}\rho U_{\cal G}^\dagger
	={}&
	\rho-i\Delta t[ H_{\cal G},\rho]
	-\frac{\Delta t^2}{2}[ H_{\cal G},[ H_{\cal G},\rho]]
	+O(\Delta t^3).
	\label{sm:unitaryexp}
\end{align}
Averaging over the topology ensemble yields
\begin{align}
	\Phi_{\Delta t}[\rho]
	={}&
	\rho-i\Delta t[\bar H,\rho]
	-\frac{\Delta t^2}{2}[\bar H,[\bar H,\rho]]
	\\
	&
	-\frac{\Delta t^2}{2}
	\mathbb E\!\left[\delta H_{\cal G},[\delta H_{\cal G},\rho]\right]
	+O(\Delta t^3),
	\label{sm:channel_exp}
\end{align}
where the cross terms vanish because $\mathbb E[\delta H_A]=0$. The last
term can be written in terms of the edge-space covariance matrix,
\begin{equation}
	C_{ef}=\mathbb E[\delta A_e\delta A_f],
	\label{sm:cov}
\end{equation}
so that
\begin{equation}
	\mathbb E\!\left[\delta H_{\cal G},[\delta H_{\cal G},\rho]\right]
	=
	J^2\sum_{e,f}C_{ef}[V_e,[V_f,\rho]].
	\label{sm:noise_double_comm}
\end{equation}

The continuous generator associated with the one-step channel is defined
through the logarithm of \eqref{sm:channel},
\begin{equation}
	\mathcal L_{\rm eff}
	=\frac{1}{\Delta t}\log\Phi_{\Delta t},
	\label{sm:log_generator}
\end{equation}
where the logarithm is understood on the branch continuously connected
to the identity as $\Delta t\to0$~\cite{Breuer2002}. Writing
\begin{equation}
	\Phi_{\Delta t}
	=I+\Delta t\mathcal A+\Delta t^2\mathcal B+O(\Delta t^3),
\end{equation}
with
\begin{equation}
	\mathcal A[\rho]=-i[\bar H,\rho],
\end{equation}
and using $\log(I+X)=X-X^2/2+O(X^3)$, we obtain
\begin{equation}
	\mathcal L_{\rm eff}
	=\mathcal A+\Delta t\left(\mathcal B-\frac12\mathcal A^2\right)+O(\Delta t^2).
\end{equation}
The purely coherent second-order contribution cancels,
\begin{equation}
	\mathcal B-\frac12\mathcal A^2
	=-\frac{1}{2}\mathbb E\!\left[\operatorname{ad}_{\delta H_{\cal G}}^2\right],
\end{equation}
where $\operatorname{ad}_X[\rho]=[X,\rho]$. Therefore
\begin{equation}
	\mathcal L_{\rm eff}[\rho]
	=
	-i[\bar H,\rho]
	-
	\frac{J^2\Delta t}{2}
	\sum_{e,f}C_{ef}[V_e,[V_f,\rho]]
	+O(\Delta t^2).
	\label{sm:effective_generator}
\end{equation}
% [MAIN-EQ] verify equation number of the effective generator in the main text
This is Eq.~(6) of the main text. The finite-step channel in
Eq.~\eqref{sm:channel} remains the exact description; the continuous
generator is the leading rapid-switching approximation. Since the
dissipative prefactor is proportional to $J^2\Delta t$, the induced
rates decrease as the switching rate $\nu=1/\Delta t$ is increased,
yielding the motional-narrowing scaling discussed in the main text.

\subsection{Covariance diagonalization and Lindblad-mode construction}

Because the covariance matrix $C$ with entries $C_{ef}={\mathbb{E}}[\delta A_e\delta A_f]$, is real, symmetric, and positive semidefinite, it admits the
spectral decomposition
\begin{equation}
	C_{ef}=\sum_\alpha c_\alpha O_{e\alpha}O_{f\alpha},
	\qquad
	c_\alpha\ge0,
	\label{sm:Cdecomp}
\end{equation}
with orthogonal $O$. Defining
\begin{equation}
	W_\alpha=\sum_eO_{e\alpha}V_e,
	\label{sm:Wdef}
\end{equation}
and substituting Eq.~\eqref{sm:Cdecomp} into
Eq.~\eqref{sm:effective_generator}, we find
\begin{equation}
	\sum_{e,f}C_{ef}[V_e,[V_f,\rho]]
	=
	\sum_\alpha c_\alpha[W_\alpha,[W_\alpha,\rho]].
	\label{sm:double_comm_modes}
\end{equation}
Since every $V_e$ and hence every $W_\alpha$ is Hermitian,
\begin{equation}
	[W,[W,\rho]]=-2\mathcal D[W]\rho,
	\qquad
	\mathcal D[W]\rho=W\rho W-\frac12\{W^2,\rho\},
	\label{sm:dissipator_identity}
\end{equation}
which brings the generator to the Gorini--Kossakowski--Sudarshan--Lindblad
form~\cite{Lindblad1976,Gorini1976}
\begin{equation}
	\mathcal L_{\rm eff}[\rho]
	=
	-i[\bar H,\rho]
	+
	\sum_\alpha\Gamma_\alpha\mathcal D[W_\alpha]\rho
	+O(\Delta t^2),
	\label{sm:lindblad_form}
\end{equation}
with
\begin{equation}
	\Gamma_\alpha=J^2\Delta t\,c_\alpha.
	\label{sm:Gamma}
\end{equation}
Only eigenvalues with $c_\alpha>0$ correspond to active
topology-induced dissipative channels. Hence the number of active
channels is
\begin{equation}
	r=\operatorname{rank}C\le M.
	\label{sm:rank}
\end{equation}
This establishes explicitly the separation used in the main text: the
mean topology fixes $\bar H$, the covariance eigenvalues fix the noise
strengths $\Gamma_\alpha$, and the covariance eigenvectors determine the
collective operators $W_\alpha$.

\subsection{Decoherence-free sectors and the covariance-geometry criterion}

Let $P$ project onto a subspace $\mathcal H_{\rm DFS}$ such that, for
every active dissipative mode,
\begin{equation}
	W_\alpha P=w_\alpha P,
	\label{sm:DFS}
\end{equation}
with real $w_\alpha$. This is the standard decoherence-free-subspace
condition for Hermitian jump operators~\cite{Zanardi1997,Lidar1998}. For
any density operator supported in the subspace, $\rho=P\rho P$, one then
has
\begin{align}
	W_\alpha\rho W_\alpha&=w_\alpha^2\rho,\\
	W_\alpha^2\rho&=w_\alpha^2\rho,\\
	\rho W_\alpha^2&=w_\alpha^2\rho,
\end{align}
so that
\begin{equation}
	\mathcal D[W_\alpha]\rho=0.
	\label{sm:DFS_diss_zero}
\end{equation}
If, in addition,
\begin{equation}
	\bar HP=P\bar HP,
	\label{sm:H_invariance}
\end{equation}
the subspace is invariant under the coherent evolution. Thus the
topology-induced dissipator vanishes while the remaining motion is
unitary inside the subspace. If $\bar H P=hP$, the subspace is
stationary.

More generally, let $|a\rangle$ and $|b\rangle$ be eigenstates of a
Hermitian noise operator $W$, with
\begin{equation}
	W|a\rangle=w_a|a\rangle,
	\qquad
	W|b\rangle=w_b|b\rangle.
\end{equation}
Then
\begin{equation}
	\mathcal D[W](|a\rangle\langle b|)
	=-\frac12(w_a-w_b)^2|a\rangle\langle b|.
	\label{sm:eigenoperator_diss}
\end{equation}
This identity shows directly why a fixed dissipative strength need not
imply a fixed decay rate for different coherences: the eigenvalues of
the noise operator, and hence its eigenspace geometry, determine which
coherences are damped.

%%%%%%%%%%%%%%%%%%%%%%%%%%%%%%%%%%%%%%%%%%%

\section{Exact dynamics of the rank-one global and community ensembles}

We now derive the finite-$\Delta t$ dynamics used in Fig.~1 of the main
text. Throughout this section $N=2n$ and $\xi=1/2$.

Define the symmetric and community-antisymmetric states
\begin{equation}
	|+\rangle=\frac{1}{\sqrt N}\sum_{j=1}^N|j\rangle,
	\qquad
	|-\rangle=\frac{1}{\sqrt N}
	\left(\sum_{j\in A}|j\rangle-\sum_{j\in B}|j\rangle\right).
	\label{sm:pm}
\end{equation}
Let $\mathcal R$ denote the $(N-2)$-dimensional subspace orthogonal to
both $|+\rangle$ and $|-\rangle$.

\subsection{Global ensemble}

For the global rank-one ensemble, all edges are determined by a single
Bernoulli variable. The graph is either empty or complete, and
\begin{equation}
	\Phi_G[\rho]
	=\frac12\rho+\frac12e^{-iJS\Delta t}\rho e^{iJS\Delta t}.
	\label{sm:global_channel}
\end{equation}
The complete-graph operator $S=\sum_{i<j}V_{ij}$ obeys
\begin{equation}
	S=(N-1)P_+-P_\perp,
	\qquad
	P_+=|+\rangle\langle+|,
	\qquad
	P_\perp=I-P_+.
	\label{sm:Sspectral}
\end{equation}
Consequently,
\begin{equation}
	S|+\rangle=(N-1)|+\rangle,
	\qquad
	S|r\rangle=-|r\rangle,
	\quad |r\rangle\in\mathcal R\oplus\mathrm{span}\{|-\rangle\}.
\end{equation}
A coherence connecting $|+\rangle$ to any state orthogonal to $|+\rangle$
is therefore multiplied in one step by
\begin{equation}
	\mu_{+\perp}^{(G)}
	=\frac12\left[1+e^{-iJN\Delta t}\right]
	=e^{-i\omega\Delta t}\cos(\omega\Delta t),
	\label{sm:mu_global}
\end{equation}
with $\omega=\frac{JN}{2}$. Writing $c=\cos(\omega\Delta t)$ and $k=t/\Delta t$, the magnitude of
this coherence decays as $|c|^k$.

For the initially localized state $|1\rangle$, use
\begin{equation}
	|1\rangle=\frac{1}{\sqrt N}|+\rangle+
	\sqrt{\frac{N-1}{N}}\,|r_1\rangle,
	\qquad
	\langle +|r_1\rangle=0.
	\label{sm:local_decomp}
\end{equation}
Applying the channel to the corresponding density operator gives
\begin{equation}
	F^{(G)}(t)
	=\frac{1+(N-1)^2}{N^2}
	+
	\frac{2(N-1)}{N^2}c^{t/\Delta t}\cos(\omega t),
	\label{sm:Fglobal}
\end{equation}
which is the expression quoted in the main text.

The inter-sector Frobenius coherence
% [MAIN-EQ] verify equation number of the coherence definition in the main text
defined in Eq.~(21) of the main text is
\begin{equation}
	\mathcal C_{+\perp}^{(G)}(t)
	=\frac{\sqrt{N-1}}{N}|c|^{t/\Delta t}.
	\label{sm:Cglobal}
\end{equation}
The total purity follows from the block structure of the state,
\begin{equation}
	\operatorname{Tr}\!\left[(\rho^{(G)}(t))^2\right]
	=
	\frac{1+(N-1)^2}{N^2}
	+
	\frac{2(N-1)}{N^2}c^{2t/\Delta t}.
	\label{sm:Pglobal}
\end{equation}

\subsection{Community ensemble}

For the community ensemble, let $S_{\rm in}$ and $S_{\rm out}$ denote
the sums of the intra- and inter-community exchange operators. Then
\begin{equation}
	S=S_{\rm in}+S_{\rm out},
	\qquad
	Q=S_{\rm in}-S_{\rm out},
	\label{sm:SQdecomp}
\end{equation}
so that
\begin{equation}
	S_{\rm in}=\frac{S+Q}{2},
	\qquad
	S_{\rm out}=\frac{S-Q}{2}.
	\label{sm:S_in_out}
\end{equation}
The two realizations of the community ensemble have Hamiltonians
$JS_{\rm in}$ and $JS_{\rm out}$, each with probability $1/2$. The
operators $S$ and $Q$ commute and satisfy
\begin{equation}
	\begin{aligned}
		S|+\rangle&=(N-1)|+\rangle,\\
		S|-\rangle&=-|-\rangle,\\
		S|r\rangle&=-|r\rangle,
	\end{aligned}
	\label{sm:Scommunity}
\end{equation}
and
\begin{equation}
	\begin{aligned}
		Q|-\rangle&=(N-1)|-\rangle,\\
		Q|+\rangle&=-|+\rangle,\\
		Q|r\rangle&=-|r\rangle,
	\end{aligned}
	\label{sm:Qcommunity}
\end{equation}
with $|r\rangle\in\mathcal R$. Thus the two Hamiltonians $JS_{\rm in}$ and $JS_{\rm out}$ have eigenvalues
\begin{center}
	\begin{tabular}{lccc}
		\toprule
		& $|+\rangle$ & $|-\rangle$ & $\mathcal R$ \\
		\midrule
		$JS_{\rm in}$ & $J(N-2)/2$ & $J(N-2)/2$ & $-J$ \\
		$JS_{\rm out}$ & $JN/2$ & $-JN/2$ & $0$ \\
		\bottomrule
	\end{tabular}
\end{center}
For a coherence between $|+\rangle$ and $|r\rangle\in\mathcal R$, both
realizations produce the same relative phase $e^{-i\omega\Delta t}$;
hence this coherence has unit-modulus channel eigenvalue. By contrast,
coherences between $|-\rangle$ and $\mathcal R$ acquire an averaged
factor proportional to $\cos(\omega\Delta t)$.

For the localized state $|1\rangle$ in community $A$, the decomposition
in the community basis reads
\begin{equation}
	|1\rangle = \frac{1}{\sqrt N}\left(|+\rangle+|-\rangle\right)
	+\sqrt{\frac{N-2}{N}}\,|r\rangle.
	\label{sm:local_decomp_community}
\end{equation}
Evaluating the channel on $|1\rangle\langle1|$ gives
\begin{equation}
	\begin{aligned}
		F^{(C)}(t)
		={}&
		\frac{(N-2)^2+2}{N^2}
		+
		\frac{2}{N^2}c^{t/\Delta t}\cos(\omega t)
		\\[2pt]
		&+
		\frac{2(N-2)}{N^2}\cos(\omega t)
		+
		\frac{2(N-2)}{N^2}c^{t/\Delta t}.
	\end{aligned}
	\label{sm:Fcommunity}
\end{equation}
The sector-resolved coherence is
\begin{equation}
	\mathcal C_{+\perp}^{(C)}(t)
	=\frac{1}{N}
	\sqrt{N-2+c^{2t/\Delta t}},
	\label{sm:Ccommunity}
\end{equation}
while the purity is
\begin{equation}
	\operatorname{Tr}\!\left[(\rho^{(C)}(t))^2\right]
	=
	\frac{1+(N-1)^2}{N^2}
	+
	\frac{2(N-1)}{N^2}c^{2t/\Delta t}.
	\label{sm:Pcommunity}
\end{equation}
Equations~\eqref{sm:Pglobal} and \eqref{sm:Pcommunity} are identical,
establishing analytically the equal-purity property highlighted in the
main text.

Using
\begin{equation}
	\ln|c|=-\frac{\omega^2\Delta t^2}{2}+O(\Delta t^4),
	\label{sm:lnc}
\end{equation}
we obtain the rapid-switching decay rate
\begin{equation}
	\gamma_{+\perp}=\frac{\omega^2\Delta t}{2}
	=\frac{J^2N^2\Delta t}{8}
	\label{sm:gamma_rankone}
\end{equation}
for the global symmetric--orthogonal coherence.

\subsection{Partial channel correlations and the rate-difference test}

We use the same notation as in the main text: $\bm1$ is the all-ones
vector in edge space, and $\bm q$ has components $q_e=+1$ for edges
joining sites within the same community and $q_e=-1$ for edges joining
different communities.

The partially correlated channel interpolations are
\begin{equation}
	\begin{aligned}
		\Phi^{(G)}(\eta)&=
		\eta\Phi_{\rm corr}^{(G)}+(1-\eta)\Phi_{\rm Gilb}, \\
		\Phi^{(C)}(\eta)&=
		\eta\Phi_{\rm corr}^{(C)}+(1-\eta)\Phi_{\rm Gilb},
	\end{aligned}
	\label{sm:mixture}
\end{equation}
with $0\le\eta\le1$ and edge probability $\xi=1/2$. Here
$\Phi_{\rm Gilb}$ denotes the channel obtained from independently
sampled Gilbert graphs~\cite{Gilbert1959}. Because the constituent
channels have the same mean topology, the covariance matrices are
\begin{align}
	C^{(G)}(\eta)&=
	\frac14\left[(1-\eta)I_M+\eta\bm1\bm1^T\right],
	\label{sm:CGeta}\\
	C^{(C)}(\eta)&=
	\frac14\left[(1-\eta)I_M+\eta\bm q\bm q^T\right].
	\label{sm:CCeta}
\end{align}
Their spectra coincide,
\begin{equation}
	c_1(\eta)=\frac{1+\eta(M-1)}{4},
	\qquad
	c_{\alpha>1}(\eta)=\frac{1-\eta}{4}.
	\label{sm:spectrum_eta}
\end{equation}
The leading eigenvectors remain $\bm1/\sqrt M$ and $\bm q/\sqrt M$ for
the global and community constructions, respectively.

The difference between the corresponding effective generators can
therefore be written
\begin{equation}
	\mathcal L^{(G)}(\eta)-\mathcal L^{(C)}(\eta)
	=
	\frac{\eta J^2\Delta t}{4}
	\left[\mathcal D[S]-\mathcal D[Q]\right]
	+O(\Delta t^2).
	\label{sm:Ldifference}
\end{equation}
Consider a residual state $|r\rangle\in\mathcal R$. From the operator
relations above,
\begin{equation}
	\mathcal D[S](|+\rangle\langle r|)
	=-\frac{N^2}{2}|+\rangle\langle r|,
	\qquad
	\mathcal D[Q](|+\rangle\langle r|)=0.
	\label{sm:Ddifference}
\end{equation}
Hence, for the $\xi=1/2$ interpolation considered here, the
leading-order rate difference is
\begin{equation}
	\Delta\Gamma_{\rm th}(\eta)
	\equiv
	\Gamma^{(G)}-\Gamma^{(C)}
	=
	\frac{\eta J^2N^2\Delta t}{8}
	+O(\Delta t^2).
	\label{sm:rate_difference}
\end{equation}
This is the quantity tested in Fig.~2(b) of the main text.

For the numerical comparison, the decay rate is extracted from the
logarithmic slope of the geometry-sensitive coherence,
\begin{equation}
	\Gamma=-\frac{d}{dt}\ln \mathcal C_r(t),
	\qquad
	\mathcal C_r(t)=|\langle+|\rho(t)|r\rangle|,
	\label{sm:rate_fit}
\end{equation}
using the same finite time window and numerical procedure as in the
figure-generation code. The relative deviation shown in Fig.~2(c) of
the main text is
\begin{equation}
	\epsilon_\Gamma(\eta)
	=
	\frac{|\Delta\Gamma_{\rm num}(\eta)-\Delta\Gamma_{\rm th}(\eta)|}
	{\Delta\Gamma_{\rm th}(\eta)},
	\qquad \eta>0.
	\label{sm:relative_error}
\end{equation}
The denominator vanishes at $\eta=0$, so that point is excluded. The
observed nonmonotonicity of $\epsilon_\Gamma$ reflects the competition
between numerical noise at small $\eta$ and finite-$\Delta t$
corrections at large $\eta$.

%%%%%%%%%%%%%%%%%%%%%%%%%%%%%%%%%%%%%%%%%%%%%

\section{Latent-variable realization of multimode correlations}

We introduce a binary latent variable $Z=\pm1$ with
\begin{equation}
	\Pr(Z=+1)=\Pr(Z=-1)=\frac12,
	\label{sm:Z}
\end{equation}
and conditional edge probabilities
\begin{equation}
	\Pr(A_e=1\mid Z)=\xi+a q_eZ.
	\label{sm:latent_prob}
\end{equation}
This is a standard latent-variable construction for correlated random
graphs~\cite{Goldenberg2010}. Because $q_eZ=\pm1$, validity of the
conditional Bernoulli probabilities requires
\begin{equation}
	0\le\xi\pm a\le1,
	\qquad\text{or equivalently}\qquad
	|a|\le\min(\xi,1-\xi).
	\label{sm:a_range}
\end{equation}
For the model used in Fig.~3 we specialize to $\xi=1/2$ and choose
$0\le a\le1/2$.

The conditional mean is
\begin{equation}
	\mathbb E[A_e\mid Z]=\xi+aq_eZ.
\end{equation}
Averaging over $Z$ gives
\begin{equation}
	\mathbb E[A_e]=\xi,
	\label{sm:latent_mean}
\end{equation}
and for $e\neq f$,
\begin{align}
	\operatorname{Cov}(A_e,A_f)
	&=
	\mathbb E\!\left[
	\mathbb E[A_e\mid Z]\mathbb E[A_f\mid Z]
	\right]-\xi^2
	\\
	&=a^2q_eq_f.
	\label{sm:latent_cov_offdiag}
\end{align}
For a Bernoulli variable with marginal probability $\xi$,
\begin{equation}
	\operatorname{Var}(A_e)=\xi(1-\xi).
	\label{sm:latent_var}
\end{equation}
Thus the full covariance matrix is
\begin{equation}
	C(a)=\left[\xi(1-\xi)-a^2\right]I_M+a^2\bm q\bm q^T.
	\label{sm:latent_C}
\end{equation}
Since $\bm q^T\bm q=M$, its eigenvalues are
\begin{equation}
	\begin{aligned}
		c_{\rm coll}(a)&=\xi(1-\xi)+a^2(M-1),\\
		c_{\rm loc}(a)&=\xi(1-\xi)-a^2,
	\end{aligned}
	\label{sm:latent_eigs}
\end{equation}
with multiplicities $1$ and $M-1$, respectively.
Introducing
\begin{equation}
	\kappa=\frac{a}{\sqrt{\xi(1-\xi)}},
	\label{sm:kappa}
\end{equation}
we obtain
\begin{equation}
	C(\kappa)=\xi(1-\xi)
	\left[(1-\kappa^2)I_M+\kappa^2\bm q\bm q^T\right],
	\label{sm:Ckappa}
\end{equation}
and
\begin{equation}
	\begin{aligned}		
		c_{\rm coll}(\kappa)&=\xi(1-\xi)[1+\kappa^2(M-1)],\\
		c_{\rm loc}(\kappa)&=\xi(1-\xi)(1-\kappa^2).
	\end{aligned}
	\label{sm:kappa_eigs}
\end{equation}
For $\xi=1/2$, the chosen path $0\le a\le1/2$ corresponds to
$0\le\kappa\le1$.

\subsection{Analytical slow mode}

The effective Liouvillian of the latent-variable model has the form
\begin{equation}
	\mathcal L_{\rm eff}
	=-i[\bar H,\cdot]
	+
	J^2\Delta t\left[
	c_{\rm coll}\mathcal D\!\left[\frac{Q}{\sqrt M}\right]
	+
	c_{\rm loc}\sum_{\mu\perp q}\mathcal D[W_\mu]
	\right],
	\label{sm:multimode_L}
\end{equation}
where the $W_\mu$ form an orthonormal basis of the edge-space subspace
orthogonal to $\bm q$. We adopt the right-eigenoperator convention for
the Liouvillian~\cite{Gravina2023}.

Define
\begin{equation}
	R_{\rm slow}=P_+-P_-.
	\label{sm:Rslow}
\end{equation}
Using
\begin{equation}
	S|+\rangle=(N-1)|+\rangle,
	\qquad
	S|-\rangle=-|-\rangle,
\end{equation}
we have
\begin{equation}
	[\bar H,R_{\rm slow}]=0.
	\label{sm:Rslow_H}
\end{equation}
Moreover, both $P_+$ and $P_-$ lie within the degenerate eigenspaces of
$Q$, so
\begin{equation}
	\mathcal D[Q]R_{\rm slow}=0.
	\label{sm:Rslow_Q}
\end{equation}

To evaluate the transverse contribution, let
\begin{equation}
	P^{(e)}_{\perp}=I_M-\frac{\bm q\bm q^T}{M}
	\label{sm:Pedge}
\end{equation}
be the edge-space projector orthogonal to $\bm q$. Completeness of the
transverse eigenvectors gives
\begin{equation}
	\sum_{\mu\perp q}
	\mathcal D[W_\mu]\rho
	=
	\sum_{e,f}P^{(e)}_{\perp,ef}
	\left[
	V_e\rho V_f
	-\frac12\{V_eV_f,\rho\}
	\right].
	\label{sm:transverse_projector}
\end{equation}
For $R_{\rm slow}$ the collective component vanishes, and equivalently
one may use the identity
\begin{equation}
	\sum_{\mu\perp q}\mathcal D[W_\mu]R_{\rm slow}
	=
	\sum_e\mathcal D[V_e]R_{\rm slow}.
	\label{sm:transverse_sum_reduce}
\end{equation}
Indeed, the omitted collective contribution is proportional to
$\mathcal D[Q]R_{\rm slow}$ and therefore vanishes.

We now evaluate the remaining sum explicitly. Since
\begin{equation}
	R_{\rm slow}=P_+-P_-,
\end{equation}
its matrix elements are
\begin{equation}
	\langle i|R_{\rm slow}|j\rangle
	=\frac{1-s_is_j}{N},
	\qquad
	s_i=\begin{cases}+1,&i\in A,\\-1,&i\in B,\end{cases}
	\label{sm:Rmatrix}
\end{equation}
so the only nonzero entries are the inter-community matrix elements,
equal to $2/N$.

For an inter-community matrix element $(i,j)$ with $s_i\neq s_j$, the
jump from the edge $e=(i,j)$ alone contributes $R_{ji}=R_{ij}$, while
the anticommutator contribution from this same edge is $-R_{ij}$. Each
of the remaining $2(N-2)$ edges incident on $i$ or $j$ contributes
$-\frac12 R_{ij}$ to the anticommutator. Summing,
\begin{equation}
	\begin{aligned}
		\left(\sum_e\mathcal D[V_e]R_{\rm slow}\right)_{ij}
		&=
		R_{ij} - R_{ij} - (N-2)R_{ij}\\
		&=
		-(N-2)R_{ij}.
	\end{aligned}
\end{equation}
For matrix elements within the same community, both sides vanish.
Hence
\begin{equation}
	\sum_{\mu\perp q}\mathcal D[W_\mu]R_{\rm slow}
	=-(N-2)R_{\rm slow}.
	\label{sm:transverse_sum}
\end{equation}
Substitution into Eq.~\eqref{sm:multimode_L} gives
\begin{equation}
	\mathcal L_{\rm eff}R_{\rm slow}
	=
	-\gamma_{\rm slow}R_{\rm slow},
	\label{sm:slow_eigenvalue}
\end{equation}
with
\begin{equation}
	\gamma_{\rm slow}
	=
	J^2\Delta t\,\xi(1-\xi)(N-2)(1-\kappa^2).
	\label{sm:slow_gamma}
\end{equation}
For $N=10$, $\xi=1/2$, $J=1$, and $\Delta t=0.01$,
\begin{equation}
	\gamma_{\rm slow}=0.02(1-\kappa^2).
	\label{sm:slow_numeric}
\end{equation}
This result establishes the decay rate of an explicit Liouvillian
eigenmode. It does not, by itself, establish that no other
nonstationary mode has a smaller decay rate. For the parameter range
shown in Fig.~3, full diagonalization of the effective Liouvillian
verifies numerically that this slow mode coincides with the smallest
positive decay rate, so that $\gamma_{\rm slow}=\gamma_{\rm gap}$ for
that parameter set. The equality is therefore used as a numerical
identification of the gap, rather than as an independent analytical
bound.

\subsection{Endpoint structure at full correlation}

At $\kappa=1$, $c_{\rm loc}=0$ and the dissipator contains only the
collective operator $Q$. The two $Q$ eigenspaces are
\begin{equation}
	\mathcal H_{N-1}=\mathrm{span}\{|-\rangle\},
	\qquad
	\mathcal H_{-}=|-\rangle^\perp,
	\label{sm:Qsectors}
\end{equation}
with dimensions $1$ and $N-1$, respectively. An operator belongs to the
kernel of $\mathcal D[Q]$ precisely when it has support within a single
$Q$ eigenspace, giving
\begin{equation}
	\dim\ker\mathcal D[Q]
	=1^2+(N-1)^2
	=1+(N-1)^2.
	\label{sm:Dkernel_dim}
\end{equation}
For $N=10$, this is $82$.

The coherent Hamiltonian is $\bar H=(J/2)S$. The common eigenspace
decomposition of $S$ and $Q$ is
\begin{equation}
	\mathcal H
	=
	\mathrm{span}\{|+\rangle\}
	\oplus
	\mathrm{span}\{|-\rangle\}
	\oplus
	\mathcal R,
	\label{sm:common_decomp}
\end{equation}
with $\dim\mathcal R=N-2$. Therefore the kernel of the full Liouvillian consists of operators
supported within each common eigenspace, with dimension
\begin{equation}
	\dim\ker\mathcal L_{\rm eff}
	=1^2+1^2+(N-2)^2
	=(N-2)^2+2.
	\label{sm:Lkernel_dim}
\end{equation}
For $N=10$ this gives $66$ stationary operators.

The remaining operators in $\ker\mathcal D[Q]$ are the coherences
between $|+\rangle$ and $\mathcal R$. Their number is
\begin{equation}
	2(N-2),
	\label{sm:osc_count}
\end{equation}
which equals $16$ for $N=10$. Since $Q$ acts as $-1$ on both sectors,
these operators are unaffected by the dissipator, while the
corresponding $S$ eigenvalues differ by $N$. Thus
\begin{equation}
	\mathcal L_{\rm eff}(|+\rangle\langle r|)
	=-i\frac{JN}{2}|+\rangle\langle r|,
\end{equation}
and its Hermitian-conjugate coherence has eigenvalue $+iJN/2$. Hence
the zero-real-part sector has dimension
\begin{equation}
	1+(N-1)^2,
	\label{sm:zeroreal_dim}
\end{equation}
containing $66$ stationary modes and $16$ nonstationary oscillatory
modes for $N=10$.

At the exact discrete level, the community endpoint has the same common
decomposition. The $|+\rangle\leftrightarrow\mathcal R$ coherences
acquire a unit-modulus phase under each of the two channel components,
while the remaining dissipative modes are fixed only when their phase
factors coincide. For the nonresonant parameters used in the main text,
the exact channel therefore has $82$ unit-modulus eigenmodes, of which
$66$ are fixed points and $16$ are nonstationary oscillatory modes.

\subsection{Exact one-step channel benchmark}

For the exact one-step channel, eigenoperators satisfy
\begin{equation}
	\Phi_{\Delta t}R_\mu=\mu_\mu R_\mu,
	\label{sm:channel_eigen}
\end{equation}
and a discrete decay rate is defined by
\begin{equation}
	\gamma_\mu^{\rm exact}
	=-\frac{1}{\Delta t}\ln|\mu_\mu|.
	\label{sm:exact_decay}
\end{equation}
For finite $\Delta t$, the exact channel is therefore the
nonperturbative benchmark for the rapid-switching generator.

For the latent-variable model with $0\le\kappa<1$, the exact channel
average is evaluated numerically by Monte-Carlo sampling of graph
realizations. For each sampled graph, one value $Z=\pm1$ is drawn with
equal probability, followed by independent conditional edge draws
satisfying Eq.~\eqref{sm:latent_prob}; the corresponding unitary
$U_A=e^{-iH_A\Delta t}$ is evaluated exactly. The numerical channel is
then diagonalized as an $N^2\times N^2$ superoperator. The Fig.~3
benchmark uses $N=10$, $\xi=1/2$, $J=1$, $\Delta t=0.01$,
$5\times10^3$ graph samples, and a fixed pseudorandom seed for
reproducibility. At $\kappa=1$, the two graph realizations are
evaluated exactly without Monte Carlo sampling.

The exact-channel gap is extracted from the eigenvalues with
$|\mu_\mu|<1$ by taking the smallest positive logarithmic decay rate.
At the endpoint, eigenvalues with $|\mu_\mu|=1$ are separated into
fixed points ($\mu_\mu=1$) and oscillatory modes
($|\mu_\mu|=1$, $\mu_\mu\neq1$). This distinction is needed to
reproduce the $66+16=82$ counting given above.

\subsection{Numerical convergence and robustness checks}

The analytical slow-mode curve used in Fig.~3 is
\begin{equation}
	\gamma_{\rm slow}^{\rm th}(\kappa)
	=J^2\Delta t\,\xi(1-\xi)(N-2)(1-\kappa^2).
	\label{sm:slow_curve}
\end{equation}
The full effective Liouvillian is constructed by diagonalizing the
covariance matrix and retaining every eigenmode with nonzero covariance
eigenvalue. The numerical dissipative gap is obtained from the full
$N^2\times N^2$ Liouvillian spectrum rather than from a restricted
ansatz for the slow eigenoperator.

For the parameter set used in Fig.~3, the numerical
effective-Liouvillian gap follows the analytical slow-mode curve
Eq.~\eqref{sm:slow_curve} to numerical precision for all sampled
$\kappa<1$. The exact-channel benchmark is subject to Monte-Carlo
sampling error because the channel average is estimated from a finite
ensemble of graph realizations. Increasing the number of graph samples
changes the benchmark curve while leaving the analytical and
effective-Liouvillian curves unchanged.

For Fig.~2, the Gilbert component is generated from independently
sampled Bernoulli edges with probability $\xi=1/2$, while the
correlated component is evaluated exactly. The coherence
$|\langle+|\rho|r\rangle|$ is propagated for the switching interval
$\Delta t=0.01$ and the decay rate is extracted from a logarithmic
linear fit over the region in which the signal remains above numerical
noise. The plotted relative error compares the extracted difference
$\Delta\Gamma$ with the leading-order expression
Eq.~\eqref{sm:rate_difference}.

\end{document}